\documentclass[preprints,article,accept,pdftex,moreauthors]{Definitions/mdpi}
\firstpage{1} 
\pubvolume{1}
\issuenum{1}
\articlenumber{0}
\pubyear{2022}
\copyrightyear{2022}
\datereceived{} 
\dateaccepted{} 
\datepublished{} 
\hreflink{https://doi.org/} 

\def\kms {km\,s$^{-1}$}
\def\ergs {\,erg\,s$^{-1}$}
\def\cms {cm$^{-3}$}
\def\myr {M$_\odot$\,yr$^{-1}$}
\def\whz {W\,Hz$^{-1}$}

\def\oiii {[O\,{\sc iii}]$\lambda 5007\AA$}
\def\n2ha {[N\,{\sc ii}]/H$\alpha$\,}
\def\s2ha {[S\,{\sc ii}]/H$\alpha$\,}

\Title{The interplay between radio AGN activity and their host galaxies}

\TitleCitation{The interplay between radio AGN activity and their host galaxies}

\Author{Guilherme S. Couto $^{1}$\orcidA{} and Thaisa Storchi-Bergmann $^{2}$\orcidB{}} 

\AuthorNames{Guilherme S. Couto and Thaisa Storchi-Bergmann}

\AuthorCitation{Couto, G.S.; Storchi-Bergmann, T.}

\address{%
$^{1}$ \quad Leibniz-Institut für Astrophysik Potsdam, An der Sternwarte 16, 14482 Potsdam, Germany; gcouto@aip.de\\
$^{2}$ \quad Universidade Federal do Rio Grande do Sul, Instituto de Fisica, Porto Alegre, RS, Brazil; thaisa@ufrgs.br}

\abstract{Radio activity in AGN (Active Galactic Nuclei) produce feedback on the host galaxy via the impact of the relativistic jets on the circumnuclear gas. Although radio jets can reach up to several times the optical radius of the host galaxy, in this review we focus on the observation of the feedback deposited locally in the central region of the host galaxies, in the form of outflows due to the jet-gas interaction. We begin by discussing how galaxy mergers and interactions are the most favored scenario for triggering radio AGN after gas accretion to the nuclear supermassive black hole and star formation enhancement in the nuclear region, observed in particular in the most luminous sources. We then discuss observational signatures of the process of jet-gas coupling, in particular the resulting outflows and their effects on the host galaxy. These include the presence of shock signatures and the detection of outflows not only along the radio jet but perpendicular to it in many sources. Although most of the studies are done via the observation of ionized gas, molecular gas is also being increasingly observed in outflow, contributing to the bulk of the mass outflow rate. Even though most radio sources present outflow kinetic powers that do not reach $1\%\,L_{bol}$, and thus do not seem to provide an immediate impact on the host galaxy, they act to heat the ISM gas, preventing star formation, slowing the galaxy mass build-up process and limiting the stellar mass growth, in a ``maintenance mode" feedback.}


\begin{document}


\section{Introduction}

The Active Galactic Nuclei (AGN) population can be represented by two main categories. In~the first category, Quasars and Seyfert galaxies, sources of high bolometric luminosities, capable of generating winds through radiation pressure due to their accretion rate close to Eddington, usually found in wide-angle outflows, are commonly distinguished as {\it radiative}-mode (or sometimes also called  {\it quasar}-mode) AGNs. In~the second category, the~so called {\it radio}-mode AGN (or {\it jet}-mode, or~{\it kinetic}-mode), the~central engine launches powerful collimated jets of relativistic particles accelerated in the inner regions of the accretion disk due to its intense magnetic fields. The~origin of the difference between these two categories is believed to happen within the accretion disk structure and internal thermodynamics, and~the resulting mass accretion rates~\citep{heckman14} (and references therein).

Radio emission is one of the most distinctive tracers of nuclear activity in galaxies. The radio jets can be extremely powerful, extending up to Mpc scales and producing strong feedback in the surrounding medium of early-type galaxy hosts in the center of galaxy clusters \citep{mcnamara12,hlavacek13,tremblay18,hardcastle20}. But do radio jets produce feedback in its host galaxies? This is the central topic of this short review of the interplay between radio-emission from AGN and its host galaxies. Fast ($\gtrsim 1000\,$\kms) H\,{\sc i} 21 cm absorption outflows observed using WSRT \citep{morganti05,morganti16} and X-ray detections related to shocks signatures due to the jet-gas interactions, e.g., observed with Chandra \citep{jetha08,thimmappa22} illustrate how multi-wavelength analysis is fundamental to properly characterize the role of radio feedback.

With the emergence of Integral Field Spectroscopy (hereafter IFS) optical instruments in the past years, such as the Gemini GMOS \citep{allington-smith02}, VLT MUSE \citep{bacon10} and GTC MEGARA \citep{carrasco18}, along with deep observations in other wavelength instruments such as ALMA \citep{wootten09} and most recently JWST, recent studies of outflows in local AGN have been able to characterize and constrains the outflow properties, resolving their kinematics and determining its extent within the host galaxies. High resolution spectra allow to extract information in the ionized and molecular gas phases, such as velocity dispersion and emission line ratios, helping understand how gas excitation works. This has been proven useful in radio-loud AGN, characterized by shock-driven outflows due to jet-gas interactions.

We begin by discussing the role of interactions in the triggering of radio activity in galaxies in Section 2, since there is plenty of evidence that the host galaxies of the most powerful radio galaxies show morphological disturbances characteristic of interactions e.g. \citep{ramos11}. In Section 3 we discuss the signatures of radio-mode feedback within galaxies, as evidenced by the disturbed ionized gas kinematics in association with radio sources e.g. \citep{jarvis19}. This association is most frequently observed in the optical, in particular in recent studies IFS of the host galaxies \citep{couto20,comeron21}. These signatures point to an important role of shocks associated to the radio jets in producing feedback in the host galaxies.

A recent development in the study of AGN feedback is the observation of enhanced velocity dispersion perpendicular to the ionization axis and radio jet, and this is discussed in Section 4 e.g. \citep{ruschel21,venturi21}. In Section 5 we discuss how the increasing sensitivity of radio antennas is allowing us to find radio jets down to very low powers, as observed in the so-called ``Red Geysers" \citep{cheung16}. These galaxies show a faint AGN with mild ionized gas winds but that extend to large distances from the nucleus. The frequent presence of a radio source at the nucleus \citep{roy21}, suggests an association of the radio source with the winds, that seem to provide a ``maintenance mode" feedback. In this kind of feedback, the impact of the outflows is mild, but enough to heat and disturb the surrounding gas precluding the formation of new stars, as also been argued to happen in the central galaxies of clusters, halting the so-called ``cooling flows" e.g. \citep{mcnamara12}.

\section{Galaxy interactions as triggers of radio AGN activity}
\label{sec:inter}

We begin the discussion of radio activity in galaxies via its triggering, and~one of the possible mechanisms are galaxy interactions and mergers. While the occurrence of minor mergers seems to be frequent in galaxies hosting low-luminosity AGNs \linebreak (L$_{AGN}$ $\le 10^{44}$\ergs), major mergers are dominant in the AGN high-mass regime SMBH masses \mbox{$\ge 10^{8}$ M$_\odot$,~\citep{treister12,menci14,storchi19}.} For~the case of AGN radio activity, these sources are usually related to early-type hosts, which makes it easier to find morphological disturbances in images, non-orbital motions in the kinematics, and~support a connection with galaxy interactions as they are found in the center of galaxy clusters~\citep{colina95,veron01,inskip10}.

Galaxy interactions as a mechanism to produce dynamical destabilization of large amounts of gas have been considered a possible trigger of the AGN feeding, driving the gas from kpc-scale distances to the nucleus of the galaxy \citep{ellison08,davies22}. In a broad scenario, a major merger of gas-rich progenitors could be the responsible to concentrate gas in the central regions of the resulting galaxy, possibly triggering a starburst event close to the nucleus, while part of the gas makes its way to the central SMBH, initiating AGN activity. This scenario has been proposed in a number of powerful AGN, suggesting a link between galaxy mergers and both starburst and AGN activities \citep{emonts06,urrutia08,bessiere17,calabro19}.

Using Gemini GMOS broad-band images of a sample of 46 2 Jy radio galaxies, \citet{ramos11} performed one of the first qualitative studies of host galaxy morphologies of radio-loud AGNs at redshifts in the range $0.05 < z < 0.7$. Peculiar morphologies including fan-like structures, tails, bridges, shells and others were observed in 85$\%$ of the sample galaxies, a considerably higher fraction than observed in quiescent elliptical galaxies \citep{dokkum05,tal09}. When dividing the sample into Weak- and Strong-Line Radio Galaxies (WLRGs and SLRGs, respectively, with~WLRGs presenting an \oiii~emission line equivalent width below 10\AA), the~authors find an incidence of disturbed morphologies in $94\%$ of the SLRGs, while WLRGs present merger signatures in only $27\%$ of the galaxies, indicating there is a correlation between the AGN power and the incidence of~mergers.

\textls[-25]{More recent studies are in agreement with these results~\citep{gordon19}. With~deep imaging observations of 30 intermediate radio power AGNs,~\citet{pierce19} have found that the most radio-powerful half of the sample displays a higher incidence of interaction signatures than the less-powerful half ($67 \pm 12 \%$ and $40 \pm 13\%$, respectively, with~corresponding radio luminosities of \mbox{$23.06 <$ log (L$_{1.4~\mathrm {GHz}}$) $< 24.0$ \whz and $22.5 < $ log (L$_{1.4~\mathrm {GHz}}$) $ < 23.06$ \whz).} Even though there is no clear correlation between L$_{1.4~\mathrm{GHz}}$ and L$_{\mathrm{[O {\sc III}]}}$ luminosities, the~same fractions are observed when separating the sample in terms of L$_{\mathrm{[O {\sc III}]}}$.}

In this context, Ultra Luminous Infrared Galaxies (ULIRGs) are also important as they support the scenario of luminous nuclear starbursts triggered by merging galaxies that evolve to become luminous AGN (as will be further explored in the next section). As shown by \citet{sanders96} using IRAS observations, the properties of ULIRGS indicate that the incidence of mergers correlate with that of AGN at the highest IR luminosities (see their Table 3). The mean projected separation of the nuclei also decreases with the increase of AGN and merger fraction, suggesting that the AGN activity is more likely to happen at latter phases of the galaxy merger due to gas migration to the central regions, which also feeds the very high star formation rates observed in ULIRGs of up to $\sim 1000\,$\myr.

\subsection*{Individual~Studies}

Clear signatures of galaxy interactions in radio-loud AGNs have been largely observed in individual studies. This is the case of B2 0648 + 27, an early-type galaxy presenting interacting signatures such as tail-like structures in low-surface brightness emission~\citep{heisler94}.~\citet{emonts06} found three main events in the galaxy's history while studying its neutral gas, optical spectroscopy, and radio continuum: a major merger that happened $\gtrsim$1.5 Gyr ago; a starburst $\sim$0.3 Gyr ago; and the AGN activity that must have started $\gtrsim$0.001 Gyr ago. The~starburst event was succeeded by the AGN activity after the removal of angular momentum from the gas  driven towards the nucleus. B2 0648 + 27 seems to be between a starburst and a normal elliptical galaxy in an evolutionary sequence triggered by galaxy interaction and where the AGN activity happens in between.

\begin{figure}[H]
\includegraphics[width=\columnwidth]{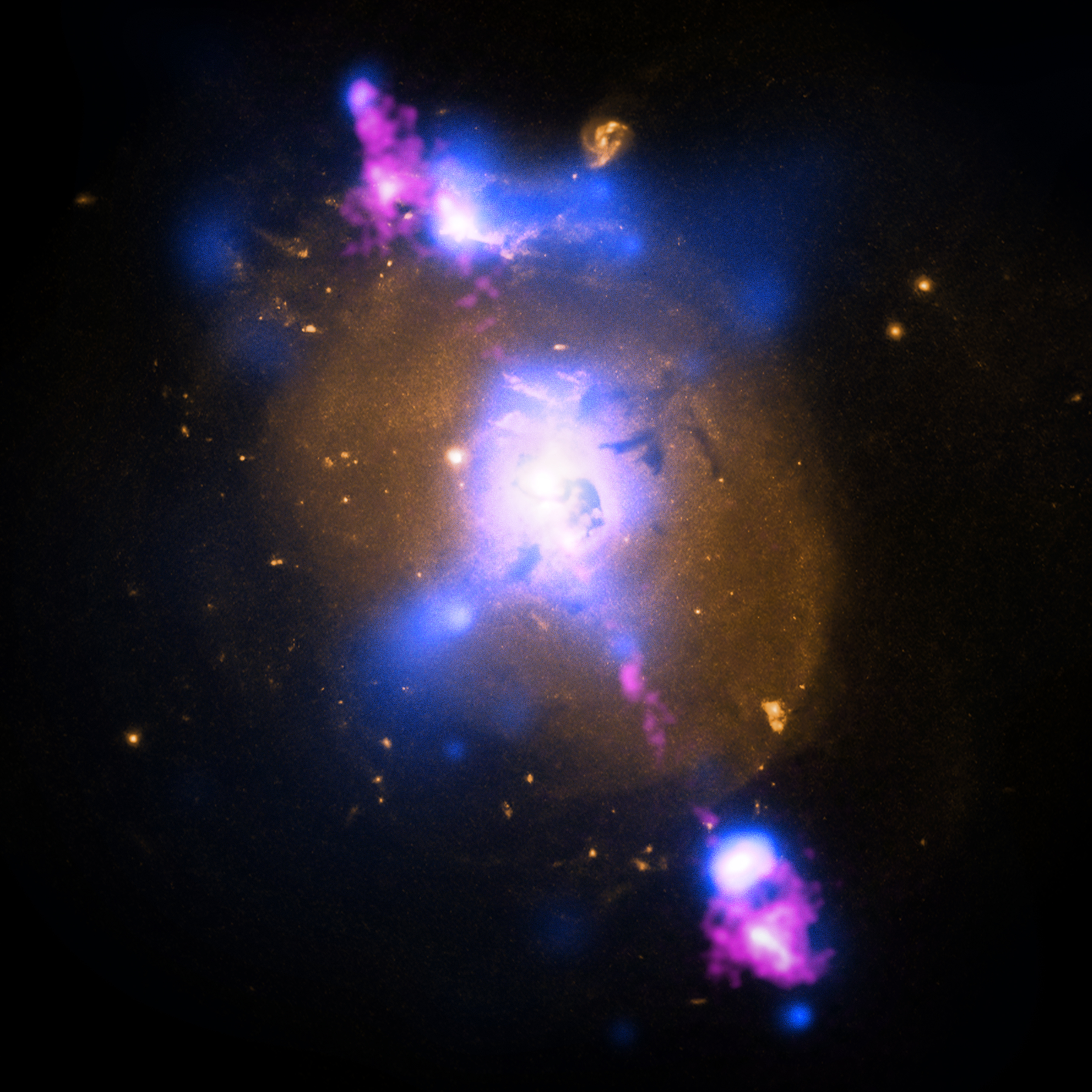}
\caption{Composite image of radio-loud AGNs 4C +29.30, with~{\it Chandra} X-ray, {\it HST} optical and VLA radio observations~\citep{siemiginowska12,sobolewska12}. Three different time periods of radio activity have been detected studying the radio-jet emission in this galaxy, which can be related to the galaxy interaction history (with clear signatures observed in the optical) and the feeding of the central~SMBH.}
\label{fig:4c29}
\end{figure}

Another interesting study case is the host galaxy of the radio-loud AGNs 4C +29.30. This galaxy seems to be a merger system in its late stages, with~a characteristic dust lane passing in front of the central region, in~a similar fashion to Centaurus A, a~well studied radio galaxy with clear signatures of past interactions~\citep{beasley08,wang20}. As~in Centaurus A, 4C +29.30 also presents several phases of recurrent radio activity, with~structures corresponding to different time periods, from~$\gtrsim$200 Myr to a very young activity of only $10^4$ yr~\citep{jamrozy07,liuzzo09}, which could relate to different feeding events during the merger process. As~shown in \mbox{Figure~\ref{fig:4c29}, 4C +29.30} shows a very complex morphology not only in the optical, but~also related to the jet in X-rays and radio wavelengths. As~we discussed in a paper detailing Gemini-GMOS IFU data~\citep{couto20}, powerful ionized gas outflows are being launched due to the interaction between the relativistic jet and the circumnuclear gas, a~scenario quite common in radio-loud AGNs that we will explore in the following sections. Several other studies on individual sources display a possible relation between radio AGN activity and galaxy interactions, which may be related ultimately to the fueling of the SMBH, or~at least driving gas to the nuclear vicinity e.g.~\citep {frank16, couto16, salome21}.

NGC 3801 is one of the few radio-loud AGNs where shock-induced X-ray emission due to jet--gas interaction was observed, and~it seems to be in a rare evolutionary stage where the enhanced star formation from a merger event has declined, and a young, powerful AGN activity is taking place~\citep{croston07,hota12}. In~this scenario, early stages of the radio source expansion such as in Fanaroff-Riley I objects~\citep{fanaroff74} have a more dramatic effect on the interstellar medium (ISM) when compared to larger scale extended sources such as FR II objects. The~decline of star formation and the role of radio feedback in heating the gas seem to be important mechanisms to drive this type of object in the `red sequence,' or~`red-and-dead' elliptical type~galaxies.

The relation between radio-loud AGNs and galaxy interactions seems to hold true also for higher redshift sources. Making use of high sensitivity {\it HST} imaging,~\citet{chiaberge15} studied the dependence of merger fractions in radio-loud and radio-quiet AGNs for redshifts $1 < z < 2.5$ when the AGN activity has peaked over cosmic history. The~authors find that $92\% \substack{+8\% \\ -14\%}$ of the powerful radio-loud galaxies ($P_{1.4~ \mathrm{GHz}}> 10^{30} $erg s$^{-1}$ Hz$^{-1}$) are merging systems, while only $38\% \substack{+16\% \\ -15\%}$ of the radio-quiet galaxies present merger~signatures. 

Another possible influence of mergers on the radio loudness of active galaxies is its impact on the spin of the SMBH. As~discussed in~\citet{chiaberge15}, if~gas accretion resulting from a merging galaxy occurs with a similar angular momentum axis as that of the SMBH, it may result in an increase in the SMBH spin velocity, while if the accretion happens in a different direction, the~SMBH spin can be slowed. These different orientations may then affect the nature of the radio-loudness of the AGN, as~it is expected that faster spinning SMBHs generate more powerful radio jets, and~the radio-loud/quiet characteristics could be explained by different accretion and SMBH spins~\citep{wilson95, tchekhovskoy10}. Of~course, this effect is expected to be more prominent at high redshifts ($z \gtrsim 2$), as~gas-rich mergers were more common in the younger Universe~\citep{volonteri13}.

The close environment of a radio-loud AGN, within~its dark matter haloes (of radius \mbox{$\sim$1 Mpc $h^{-1}$)} is denser than that of radio-quiet galaxies overall, but~this also happens when the galaxies are matched by stellar mass and redshift~\citep{donoso10,wylezalek13}. The fact that galaxies with high stellar masses do not necessarily host a radio-loud AGN indicates that the presence of radio-loud jets could be influenced by the environment of the host galaxy~\citep{hatch14}. In response, AGN radio jets seem to be important to control the star formation within the dark matter halo of its host galaxy, resulting in a slower stellar mass build-up process, and~the hosts would have lower stellar masses if the radio-mode feedback were not present~\citep{izquierdo18}.


\section{Radio-Mode Feedback in~Galaxies}
\label{sec:kin}

As pointed out in the Introduction, radio jets from giant elliptical galaxies at the center of clusters can extend up to Mpc scales and produce strong feedback in the surrounding medium, e.g.~\citep{mcnamara12,hardcastle20}. However, as this topic is being discussed elsewhere in this volume, we focus here on the effect of radio-mode feedback inside galaxies. In~this section, we present some of the characteristic features of the radio-mode feedback and how the interactions between the relativistic jets and the gas in its path manifest themselves within active~galaxies.

\subsection{Complex Gas~Kinematics}

The relativistic jets present in a radio-loud AGN, when coupled with the circumnuclear gas either in the narrow-line region (within the radius of influence of the AGN ionizing radiation) or further out in the ISM (within the inner few kpcs), can be responsible for the heating and acceleration of this gas resulting in outflows. These outflows are usually detected in the turbulent ionized emitting gas by measuring its kinematics and isolating the kinematic components linked to the AGN feedback, e.g.~\citep{cecil88,lena15,mahony16}. The~superposition of one or more kinematic components to that originating in gas rotating in the galaxy potential can result in a complex ionized gas spectrum, with~many components, making the decomposition process hard to constrain~\citep{wylezalek20}. With~the advent of better spatial and spectral resolutions IFUs, the~number of resolvable kinematic components has increased, and~the interpretation of these components and their origin has also increased in~complexity.

The case of the Seyfert 2 galaxy NGC 7130 is a clear example of such complexity, as~discussed in~\citet{comeron21}, using MUSE narrow-field adaptive optics observations. Even though the radio jet in this Seyfert galaxy is not powerful, one or maybe two outflowing ionized gas components seem to be interacting with the radio jet, as~revealed by the very detailed emission-line decomposition performed by the authors, which comprises a total of nine components, with~six being connected to outflows. It is important to note that the fitting of several components should be statistically justified so that the decomposition does not introduce artificial components into the emission-line~fit.

\begin{figure}[H]
\hspace{-0.3cm}
\includegraphics[width=12cm]{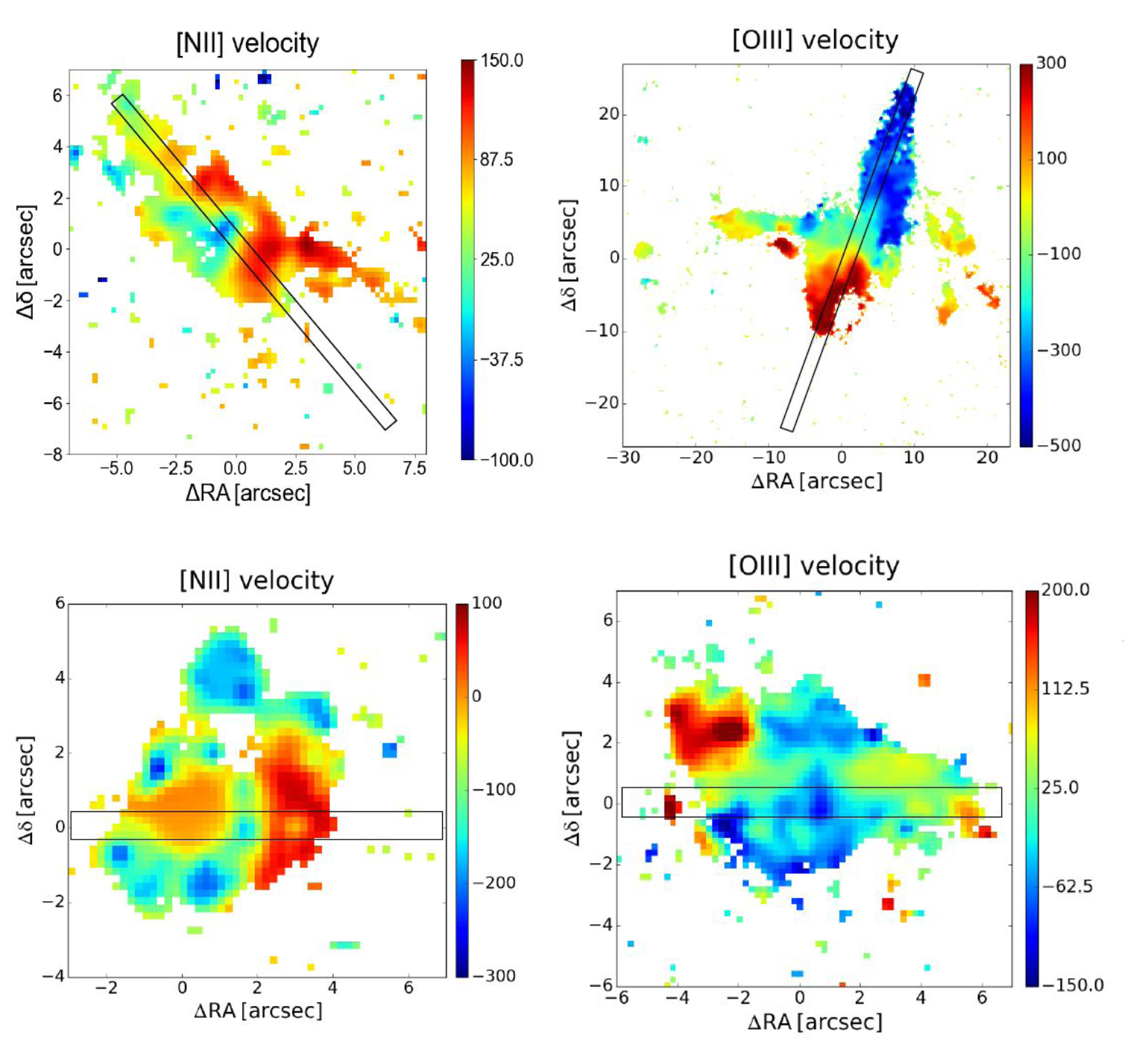}
\caption{Velocity maps in the indicated emission-lines of the radio-loud AGNs 3C 076, 3C 079, 3C 018 and 3C 017 clockwise from top left,~\citep{balmaverde19,balmaverde21}. Although some rotation seems to be present, disturbed kinematics dominate the velocity field of these galaxies, attributed to the interaction of the ambient gas with the radio jet. Velocity units are \kms.}
\label{fig:mura}
\end{figure}

Usually, a rotation component is observed in the gas velocity field of radio-loud AGN, but alongside more complex kinematics, as~illustrated in the study performed in the MURALES survey~\citep{balmaverde19,balmaverde21}, in~which a sample of 37 radio galaxies from the Third Cambridge Catalog (3C) were observed with the MUSE IFS. As shown in Figure~\ref{fig:mura}, the~ionized gas velocity maps of these radio galaxies appear to show some rotation, but~present a much more disturbed pattern than the expected ``web diagram'' characteristic of the isovelocity curves of ordered rotation. Although~these complex kinematics may not be the case for all radio-loud AGNs, they are commonly observed, as~several other studies of individual galaxies have also shown, such as in 4C +29.30~\citep{couto20}, Cygnus A~\citep{riffel21}, and UGC 05771~\citep{zovaro19b}, and~are usually caused by the interaction between the radio jet and the circumnuclear~gas. 

\begin{figure}[H]
\includegraphics[width=\textwidth]{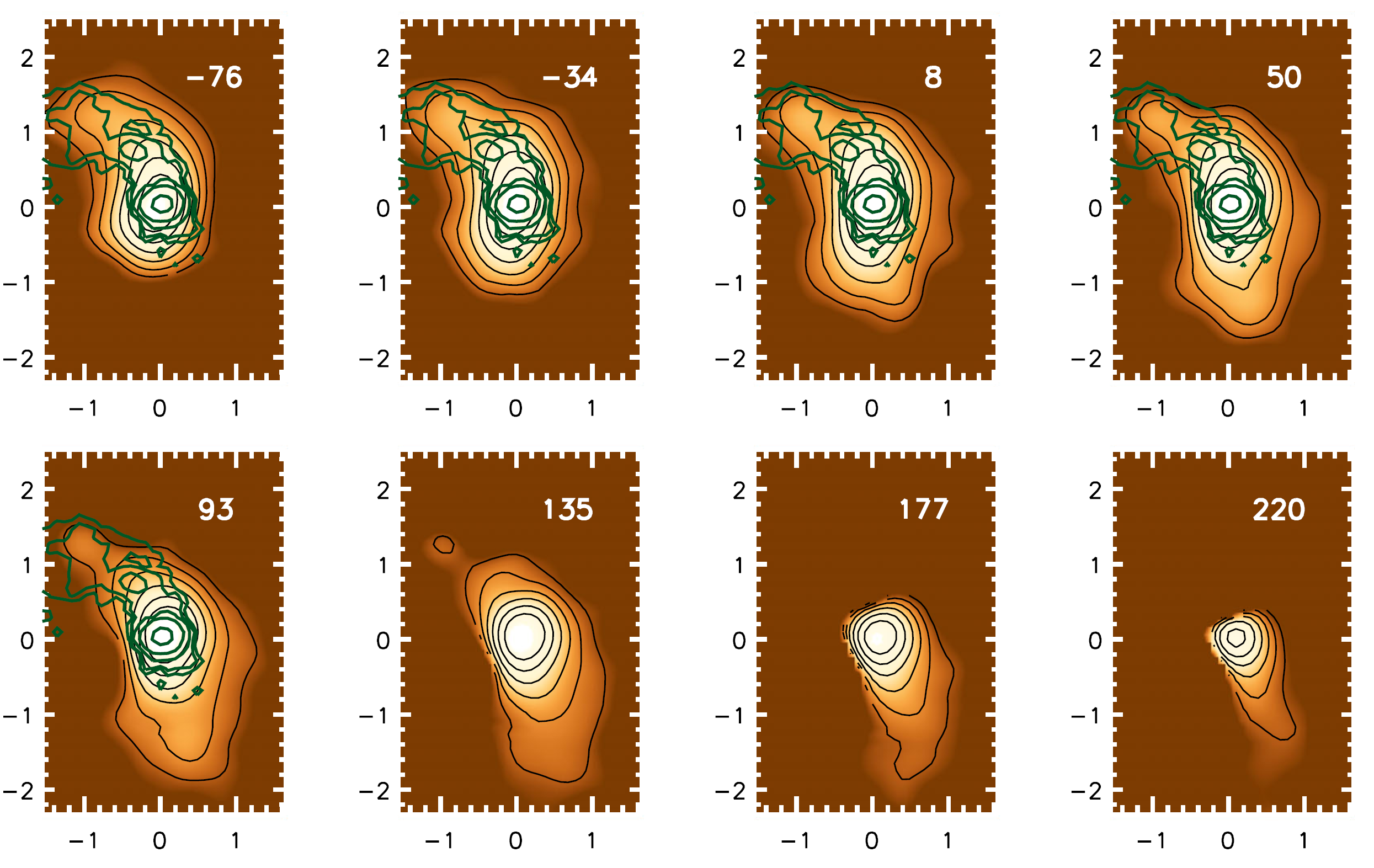}
\caption{Channel maps along the H$\alpha$ emission-line profile of the radio galaxy Arp 102B, with~velocities displayed in white at the top right corner of each map in units of \kms,~\citep{couto13}. The~flux distribution maps show a spatial correlation with the radio jet, represented by the green contours from VLA observations at 8.4 GHz. We have interpreted this emission to be related to outflowing gas pushed aside by the jet moving close to the sky plane, thus displaying blueshifted and redshifted velocities originating in the gas surrounding the radio jet. X and Y-axis are in arcsec and are centered at the galaxy~nucleus.}
\label{fig:arp}
\end{figure}

\textls[-5]{An alternative method for tracking different kinematic components is to probe the velocity field along channel maps, which is possible when using IFS observations. Arp 102B is one case in which the channel maps can aid in the interpretation of how the radio jet interacts with the surrounding emitting gas. In~\citet{couto13}, we have used Gemini-GMOS IFU observations to analyze the ionized gas kinematics in this galaxy. H$\alpha$ channel maps, shown in Figure~\ref{fig:arp}, indicate that a spiral arm-like structure correlates spatially with the radio jet, and~the emitting gas is observed both in blueshifted and redshifted velocities. We have interpreted that these velocities trace the outflowing gas pushed by the radio jet oriented very close to the plane of the sky. The~channel maps show emission from the ``walls'' surrounding the outflow being pushed aside, seen in both blueshifted and redshifted velocities. The~use of channel maps to interpret the relationship between the jet and the gas in individual galaxies studies has been proven useful in the past years, as~displayed in other papers by our group, such as in~\citet{schnorr14,lena15,riffel15}.}

\unskip

\subsection{Radio~Bubbles}

When the jet collimation is lost due to interaction with dense gas or jet precession, sometimes a gas bubble is formed, and~the inflation of the bubble by the outflow also impacts the ISM. A~nuclear starburst leading to numerous supernovae explosions can also produce outflowing bubbles from galaxy centers. The~bubble serves as a shock front between the outflow and the ISM gas, and~when it erupts, several galaxy properties can be altered, including its chemical composition. Perhaps the most famous case is of our own galaxy. Bipolar bubbles at the Galaxy center have been observed in several wavelengths, including in the radio at 1.3 GHz with MeerKAT~\citep{heywood19}, and~it is still debatable whether these bubbles originate from AGNs or star formation feedback. A~similar case where there is ambiguity in the feedback origin is NGC 3079~\citep{cecil01}. While studying the AGN-starburst composite galaxy NGC 6764,~\citet{hota06} also found an inconclusive origin for the feedback, and~compared it to 10 other sources. These are all associated with AGN activity, but the radio and optical emissions could be also affected by central starbursts. Pure starburst sources do not seem to present the same structures, indicating that AGNs must be present to produce the large-scale~bubbles.

\subsection{Molecular~Gas}
Besides the effect on the ionized gas phase that traces the hot and turbulent outflows, molecular gas can also show signatures of jet--gas interaction. As~the ionized gas represents only a fraction of the total gas mass, which is dominated by the molecular gas in the galaxy's inner regions, the~outflow gas mass should also be dominated by the molecular gas, especially in low AGN bolometric luminosities~\citep{carniani15}. Mass outflow rates observed in molecular gas are about 2--3 orders of magnitude higher than those traced in the ionized gas phase ($\sim$1000 \myr as compared to $\sim$1--10 \myr, respectively) in  AGNs with bolometric luminosity $\sim$$10^{46}$ \ergs e.g.~\citep{fiore17}. As~a comparison, star-formation rates of \linebreak 10--100 \myr (see Figure~3 in~\citet{fiore17}) are estimated for AGNs in the same luminosity range demonstrating that these outflows must indeed originate from AGN~activity.

As shown in a study of the galaxy IC 5063 by~\citet{dasyra22}, by~estimating the internal and external pressures of the molecular clouds, one can infer the impact of the radio jet in the star formation processes within the galaxy, where both suppression and enhancement can simultaneously happen. In~NGC 613, disturbed gas can be traced in the nucleus due to molecular outflows mainly boosted by the radio jet~\citep{audibert19}. While modeling the molecular gas outflows in the young radio galaxy 4C 31.04,~\citet{zovaro19a} could reproduce the observed kinematics by assuming that the gas was being pushed and expanded by the radio jet in an energy bubble while generating shocks within this bubble, originating the observed H$_2$ emission. As~observed (and discussed above) for the ionized gas phase, rotation is also usually observed in the molecular gas phase but~with distortions commonly found in the inner few hundred pc, as~analyzed by~\citet{ruffa19} in a sample of six low excitation radio galaxies using ALMA observations. However, in~the case of this sample, the~authors interpreted that these non-rotating components were related to inflowing gas since they seem to be correlated with structures such as spirals or bars, known for being mechanisms causing the gas to lose angular momentum to reach the nucleus and feed the SMBH. This illustrates that other signatures, besides the gas velocities should be used in the search for outflows, such as velocity dispersion and line ratios and their relation to the observed kinematics. Other cases of galaxies presenting jet--gas interaction signatures through the analysis of their molecular gas kinematics include NGC 1377~\citep{aalto16}, ESO 420-G13~\citep{fernandez20}, and NGC 7319~\citep{pereira22}, among~others.

\subsection{Models and~Simulations}
\begin{figure}[H]
\centering
\includegraphics[width=\textwidth]{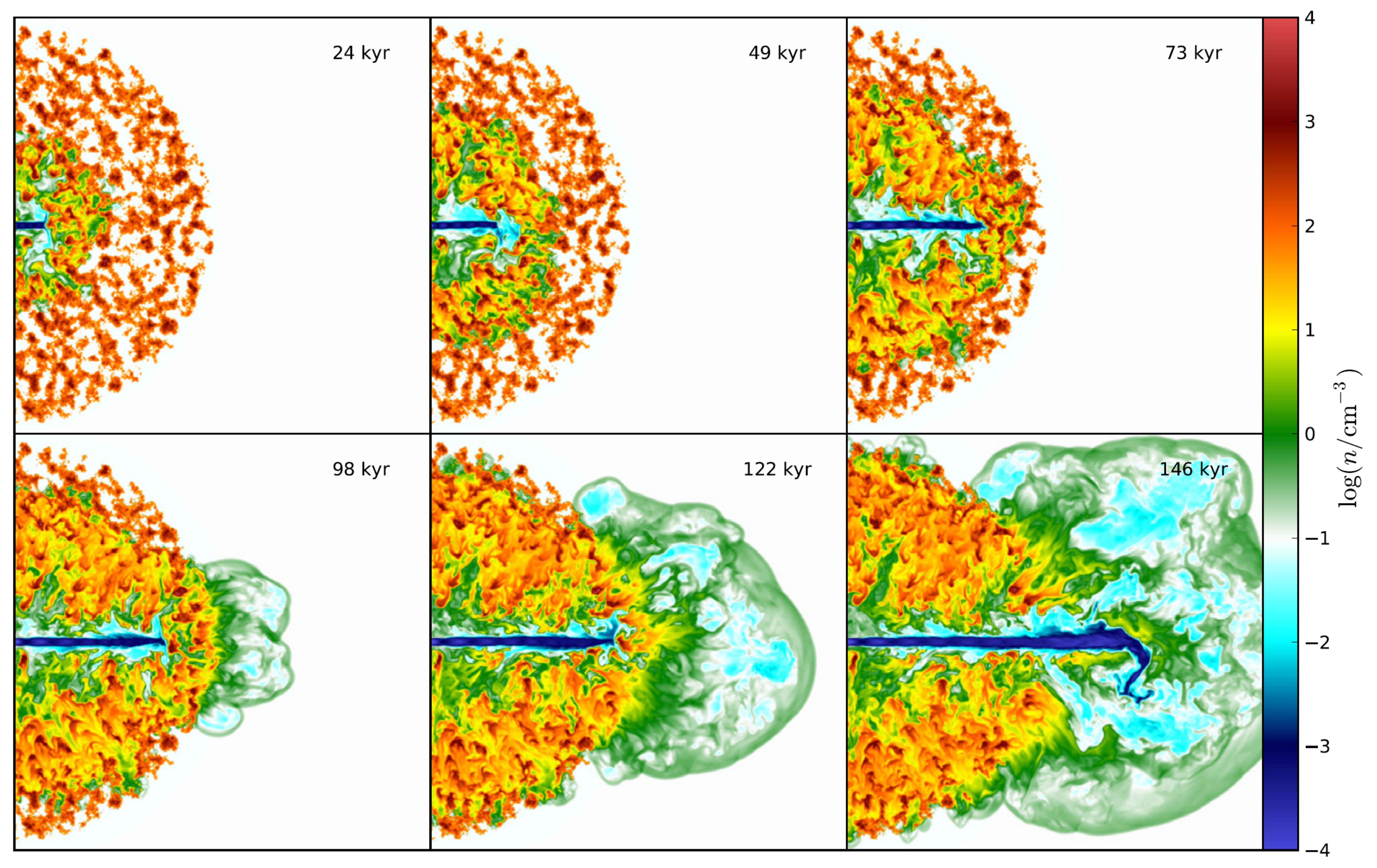}
\caption{Figure 2 from~\citet{wagner11}, showing the time evolution of the jet--gas interaction in one of the simulations performed in the paper. Colors represent the different gas densities as the jet makes its way through it, being deflected towards regions of less resistance and finally creating a bubble of expanding gas. Some of the dispersed gas may reach velocities of up to $\sim$$1000~$\kms.}
\label{fig:wag}
\end{figure}

Detailed 3D hydro-dynamical simulations indicate that, indeed the gas can be perturbed by the radio jet within the host galaxy. Not only is the jet responsible for disturbing and shaping the emitting gas distribution, as~discussed in~\citet{wagner11}, but the~path taken by the jet and its morphology is also affected by the inhomogeneity of the gas density. In this scenario, the~jet collimation, its power, and how it spatially couples with the gas are important parameters to estimate the feedback energetics. One example of such simulations is displayed in Figure~\ref{fig:wag}, where the gas density distribution is shown while the radio jet evolves with time: the jet carves its way through the gas, pushing it both aside and forward to larger distances while heating it and possibly creating shock ionization. How easy and straight the path of the jet is through the gas depends on its density and porosity distributions. We refer the reader interested in learning more about simulations of jet--gas interactions to papers by~\citet{mukherjee18a,mukherjee18b,talbot21,meenakshi22} for further~details.

\unskip

\subsection{Signatures of Shocks Due to Radio~Jets}
\label{sec:shock}

During the interaction between the relativistic jet and the surrounding gas, gas ionization through fast shocks can occur, with~the excited gas emitting characteristic spectra that provide information about the physical parameters of the shock ionization. Observed line ratios in the narrow-line region (NLR) indicating the presence of shocks usually lie in the Low-Ionization Emission Line Region LINER, ~\citep{heckman80} part of optical diagnostic diagrams, such as the well-known BPT diagrams~\citep{baldwin81}, depending on parameters such as the shock velocity and gas density~\citep{dopita95,allen08}. Although~photoionization models have a considerable overlap with shock models in the BPT diagrams, high values of low-ionization line ratios such as \n2ha and \s2ha are usually interpreted as tracers of shocks when gas emission is observed due to jet~influence. 

The presence of shocked gas also commonly correlates with broader emission lines, which trace the increase of gas turbulence. This is the case of 3C 293~\citep{mahony16}, where jet-driven outflows are detected along the radio emission, and two kinematic components are needed to reproduce the ionized gas emission-line profiles. A~clear increase in the \n2ha and \s2ha line ratios is observed in the regions where a broader component is required to reproduce the emission lines [see Figure~6 in~\citep{mahony16}], indicating that the shock velocity increases in the regions where the highest ratios are~observed.

\begin{figure}[H]
\hspace{-0.2cm}
\includegraphics[width=0.8\textwidth]{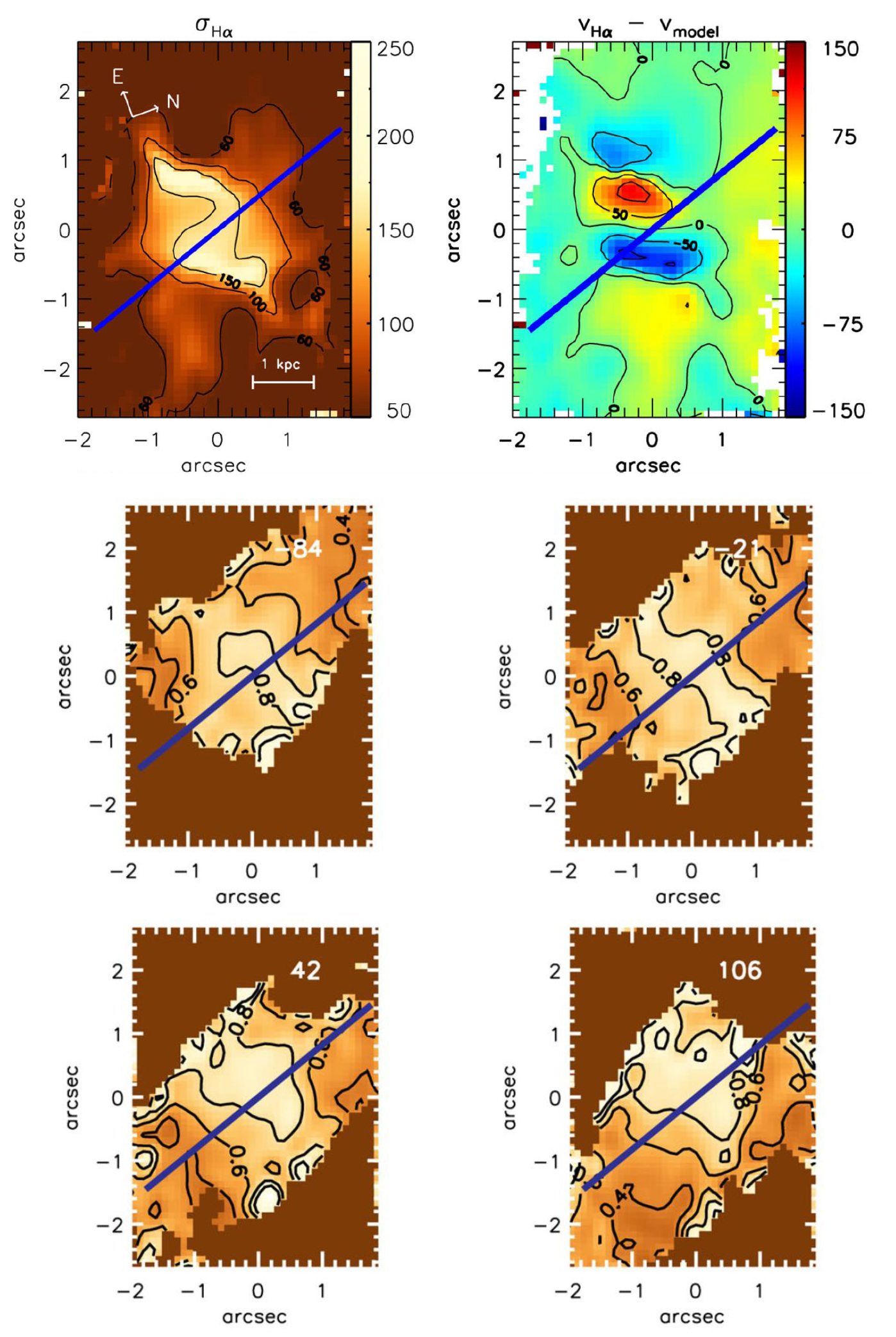}
\caption{Gemini-GMOS IFS observations of the radio galaxy 3C 33. H$\alpha$ velocity dispersion (top left), velocity residuals (top right, observed velocity field minus disk-like rotation model) and \n2ha line ratio channel maps (middle and bottom panels) adapted from~\citet{couto17}. Channel map velocities ($-84$, $-21$, 42 and 106 \kms, clockwise from the top left) are displayed in the top right corner of the panels. The~blue line displays the large-scale radio jet orientation. An~increase in the velocity dispersion is observed in the same region where velocity residuals are highest. The~same velocities are spatially correlated with an increase of the \n2ha line ratio (delineated by contours), indicating that shock-driven outflows are present in these regions, probably due to the jet--gas~interaction.}
\label{fig:3c33}
\end{figure}

Another interesting case is the gas excitation observed in 3C 33~\citep{couto17}. As~displayed in Figure~\ref{fig:3c33}, velocity dispersion increases in a region surrounding the nucleus, perpendicular to the radio jet (a discussion that we will explore further in Section~\ref{sec:perp}), in~the same orientation where high-velocity residuals are observed when a rotating disk model is subtracted from the H$\alpha$ velocity field. This correlation indicates that this region is dominated by a non-rotating kinematic component. Using channel maps of the \n2ha line ratio, we could detect the increase of this line ratio with the velocity residuals since the regions presenting the highest blueshifted and redshifted residuals show high \n2ha values with the same velocity shift. Other shock-driven outflow signatures, such as an increase of gas temperature (obtained using the [O {\sc iii}]$\lambda 4959 + 5007/4363$ line ratio) and density, strengthen the scenario that the excitation of this component is (at least partially) due to kinetic-mode~feedback.

Shock excitation has been found in several other objects presenting radio jets, such as M51~\citep{cecil88}, Coma A~\citep{morganti02}, the~Beetle galaxy~\citep{villar17}, PKS B1934-63~\citep{santoro18}, 3C 320~\citep{vagshette19}, 4C 31.04~\citep{zovaro19a}, 3C 433~\citep{murthy20}, J1220 + 3020~\citep{molina21}, and Cygnus A~\citep{riffel21}, among~others. Shock excitation has been found not only in the ionized gas phase but~also sometimes in molecular or neutral gas. High gas velocities of up to $\sim$1000 \kms, or~even higher, are commonly found in these sources, and~an increase of velocity dispersion is also a good tracer of the regions presenting~shocks.

\subsection{Feedback Power and Scaling~Relations}
\label{sec:scal}

After characterizing gas outflows due to the AGN jet feedback, it is useful to estimate its kinetic power in order to quantify its impact on the host galaxy and compare it with models and other sources of feedback. The~kinetic power of the outflow attributed to the kinematic disturbance produced by the radio jet can be calculated via~\citep{holt06}:
\begin{equation}
\dot{E} =  6.34 \times 10^{35} \frac{\dot{M}_{out}}{2} (v^2_{out} + 3\sigma^2)  ,
\end{equation} 

\noindent
where $\dot{M}_{out}$ is the mass outflow rate, $v_{out}$ is the deprojected outflow velocity, and $\sigma$ is the outflow velocity dispersion. The~mass outflow rate is dependent on the assumed outflow geometry, and~can be expressed through:
\begin{equation}
\dot{M} = 1.4  n_e  m_p  v_{out}  A  f  ,
\end{equation} 

\noindent
where $n_e$ is the gas density, $A$ is the assumed geometric cross-section area of the outflow, $f$ is the filling factor within the outflow volume, and $m_p$ is the proton mass ($m_p = 1.7 \times 10^{-24} $g), while the factor 1.4 accounts for the contribution of elements heavier than~hydrogen.

The calculations above are not restricted to radio-loud AGNs but~are general to any type of outflow. For~the ionized gas, the~gas density is usually estimated using \linebreak the [S {\sc II}]$\lambda6717,31\AA$ line ratio, and~the filling factor can be obtained using a measured hydrogen emission-line luminosity, such as H$\alpha$~\citep{peterson97}:
\begin{equation}
f = 2.6 \times 10^{59} \frac{L_{41}(\textrm{H}\alpha)}{V  n_3^2}  ,
\end{equation} 

\noindent
where $L_{41}(\textrm{H}\alpha)$ is the H$\alpha$ luminosity in units of $10^{41} $\ergs, $V$ is the assumed geometry volume, and~$n_3$ is the gas density in units of $10^3 $\cms.

Scaling relations using the estimated mass outflow rates $\dot{M}$ and outflow kinetic power $\dot{E}$ of AGN-driven feedback have been the object of studies in the past few years.~\citet{fiore17} investigated the relation between the bolometric luminosity and both the mass outflow rate and outflow kinetic power for 94 AGNs covering a large range of redshifts (from local Universe up to z$\sim$6) compiling observational data from the literature and homogeneously calculating the mass outflow rates and powers. Ionized and molecular gas show clear correlations between $L_{bol}$ and both $\dot{M}$ and $\dot{E}$, with~somewhat higher $\dot{M}$ and $\dot{E}$ for the molecular gas that seems to reach the coupling efficiency in the range \mbox{1--10\% $L_{bol}$,} as~required in models e.g.~\citep{dimatteo05,hopkins10} to have a significant impact on the host galaxy (e.g.,~by~pushing the gas out of the galaxy and halting star-formation) while for the ionized gas only $\approx$30\% of the galaxies reach this~efficiency.

\textls[-15]{More recent studies, e.g.~\citep{baron19,dallagnol21} and references therein on kinematic feedback via ionized gas outflows have obtained higher gas densities and resulting lower powers than those estimated in~\citet{fiore17} for most AGNs. In~some cases, like in the radio-galaxy \mbox{4C +29.30~\citep{couto20},} the~outflow power can reach a few percent of the AGN power $L_{AGN}$, implying strong feedback, but in~most cases, this power is below 1\%$L_{AGN}$. The~kinematic feedback is usually present, heating and disturbing the gas---a ``maintenance mode feedback,''  but~not high enough to push the gas out of the galaxy or immediately halt star formation. On the other hand, AGN feedback occurs not only via outflows, with~recent model estimates suggesting that at most 20\%\ of the AGN feedback is in kinetic form, e.g.~\citep{richings18}.}

The relations between $\dot{E}$ and $L_{AGN}$ and $\dot{M}$ and $L_{AGN}$ discussed above seem to apply both to radio-loud and radio-quiet AGN. In~the case of the latter,~\citet{villar21} have shown that, for~a sample of 13 nearby ($z < 0.2$) radio quiet QSOs, most of which showed signatures of interactions, 10 had extended radio emission. In~addition, they found this radio emission was correlated with the optical H$\alpha$ emission, indicating jet--gas interaction. Thus radio-mode feedback is also present in radio-quiet AGNs when the jet is spatially coupled with the ISM~gas. 

\section{Statistical Studies and Higher Redshift~Sources}\label{section4}
There are a number of statistical studies of an AGN feedback effect in general (not restricted to radio AGN) on host galaxy properties.~\citet{wylezalek16} gathered a large sample of radio-quiet AGNs at $z < 1$ with SDSS observations to investigate the relation between AGN feedback and star formation quenching. Measuring the \mbox{\oiii  velocity} width as an outflow tracer, the~authors find no correlation between mass outflow rates and star formation rates; however, in~galaxies with high specific star formation rates (sSFR), they found a negative correlation between outflow strength and sSFR, indicating that in these galaxies with~the highest gas content, there is quenching due to AGN~outflows. 

Studying a much larger sample of SDSS AGN,~\citet{mullaney13} performed a multiple-component fit to the optical emission lines and found that AGNs with moderate radio luminosities (L$_{1.4~ \textrm{GHz}}=10^{123}-10^{25}$ W Hz$^{-1}$) present the most disturbed gas kinematics with the highest incidence of extremely broad \oiii  emission. This suggests that young and compact radio sources seem to more effectively drive gas turbulence in the host galaxies than powerful extended jets of the most radio luminous~sources. 

At higher redshifts, we cite the work of~\citet{delvecchio17}, where they performed a multi-wavelength analysis of 7700 radio selected AGNs from the COSMOS field with VLA observations with redshifts up to $z \lesssim 6$. They divided the sample into moderate-to-high and low-to-moderate radiative luminosity AGNs (MLAGNs and HLAGNs, respectively). The~authors found that the HLAGNs had systematically higher radiative luminosities and the AGN power occurred predominantly in radiative form, while for MLAGN, the AGN power also had a large mechanical component. They also found that at $z<1.5$, MLAGNs resided in more massive and less star-forming galaxies compared to HLAGN. At~$z>1.5$, the~opposite was found: the HLAGNs seemed to occur in more massive galaxies. The~authors interpreted their findings as evidence of downsizing, with~the most massive galaxies triggering AGNs earlier than the less massive galaxies. At~lower redshifts, the~HLAGNs faded to~MLAGNs. 


\section{Feedback Perpendicular to the Radio Jet~Orientation}
\label{sec:perp}

Even though gas outflows are usually expected to be found along the AGN radio jet, this is not always the case. Recent studies have been showing different orientations between the radio jet axis and the direction of an observed increase of the gas velocity dispersion related to outflowing gas. In~many cases, the direction of the enhanced velocity dispersion is approximately perpendicular to that of the radio jet. Perhaps the first observational evidence of such orientation in an IFS analysis was the case of Arp 102B~\citep{couto13}. In~this radio galaxy, we observed an increase in the gas velocity dispersion in the [O {\sc iii}]$\lambda 5007 \AA$ emission line, close to the nucleus, approximately perpendicular to the radio jet. The~jet then bends with increasing distance to the~nucleus.

\begin{figure}[H]
\includegraphics[width=\textwidth]{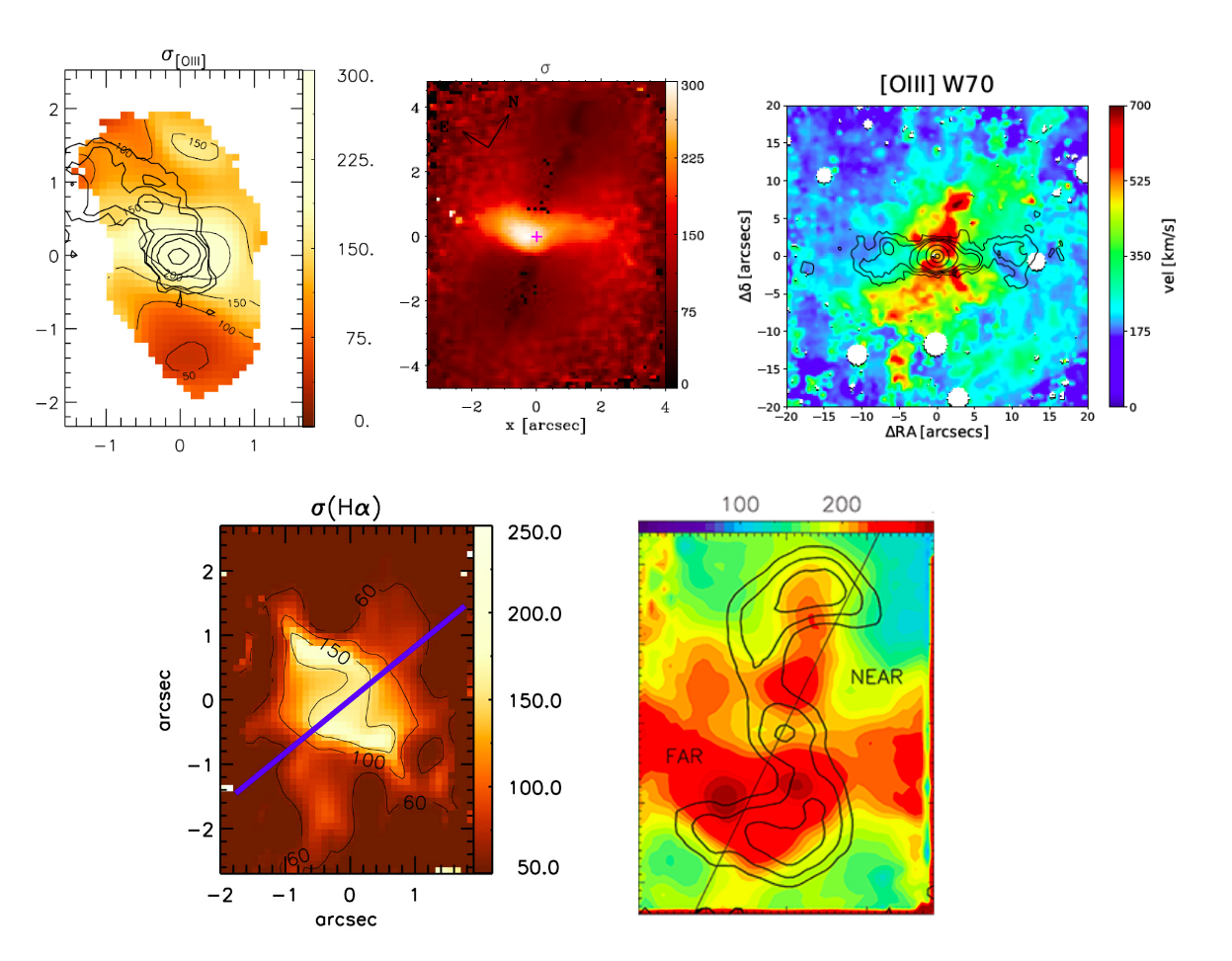}
\caption{Gas kinematics of several active galaxies displaying increased gas velocity dispersion -- attributed to outflowing components -- perpendicularly (or close to being perpendicular) to the radio jet orientation. Top left: Arp 102B Gemini GMOS-IFU [O {\sc iii}] velocity dispersion map, with~VLA 8.4~GHz contours displaying the radio jet emission~\citep{couto13}. Top middle: NGC 1386 GMOS-IFU [N {\sc ii}] velocity dispersion map, with~the dark region along the north--south axis tracing the jet emission (not shown here) and enhanced integrated flux~\citep{lena15}. Top right: NGC 5643 VLT-MUSE [O {\sc iii}] W70 map showing highest values perpendicular to the radio jet shown via VLA 8.4 GHz emission contours~\citep{venturi21}. Bottom left: 3C 33 H$\alpha$ velocity dispersion map, with~the large scale VLA 1.4 GHz radio jet orientation represented by the blue line~\citep{couto17}. Bottom right: NGC 3393 GMOS-IFU [N {\sc ii}] velocity dispersion map also displaying VLA 8.4 GHz radio emission as black contours~\citep{finlez18}. Velocity dispersion and W70 units are in \kms.}
\label{fig:perp}
\end{figure}

Ionized gas velocity dispersion maps of Arp 102B and other galaxies presenting enhanced values approximately perpendicular to radio jets are shown in Figure~\ref{fig:perp}. Arp 102B, NGC 1386~\citep{lena15}, NGC 5643~\citep{venturi21}, 3C 33~\citep{couto17} and NGC 3393~\citep{finlez18} show very clear signatures of the misalignment between the radio jet and the extended high-velocity dispersion emission, very close to perpendicularity. In~a very detailed study of this phenomenon,~\citet{venturi21} used VLT MUSE observations to trace the outflowing gas in four galaxies displaying a mismatch relative to the orientation of the radio jets, including NGC 5643, shown in Figure~\ref{fig:perp}. Extended emission ($\gtrsim$1 kpc) is observed with increased velocity dispersion (W70 $\gtrsim$ 800--1000 \kms) perpendicularly to the jet. The~authors also found that the gas excitation in these regions was consistent with shock ionization, similar to the results we have found in 3C 33 (as discussed in Section~\ref{sec: shock}, where higher shock-related line ratio values ([S {\sc ii}]/H$\alpha$ in their case) are observed. Further evidence of this phenomenon was observed in the AGNIFS survey~\citep{ruschel21}, comprising 30 local AGNs observed with GMOS-IFU, in which outflows perpendicular to the ionization axis (via enhanced W80 values) were detected in 7 sources out of 21 that presented outflows. The~same was found in the MURALES survey~\citep{balmaverde19}, where at least 6 of the initial 20 observed galaxies present some indication of turbulent ionized gas perpendicular to the radio~jet. NGC 5929 \citep{riffel15} is another case where the velocity dispersion of the ionized gas is found to be enhanced perpendicularly to the radio jet (see their Fig. 8).

\begin{figure}[H]
\includegraphics[width=\textwidth]{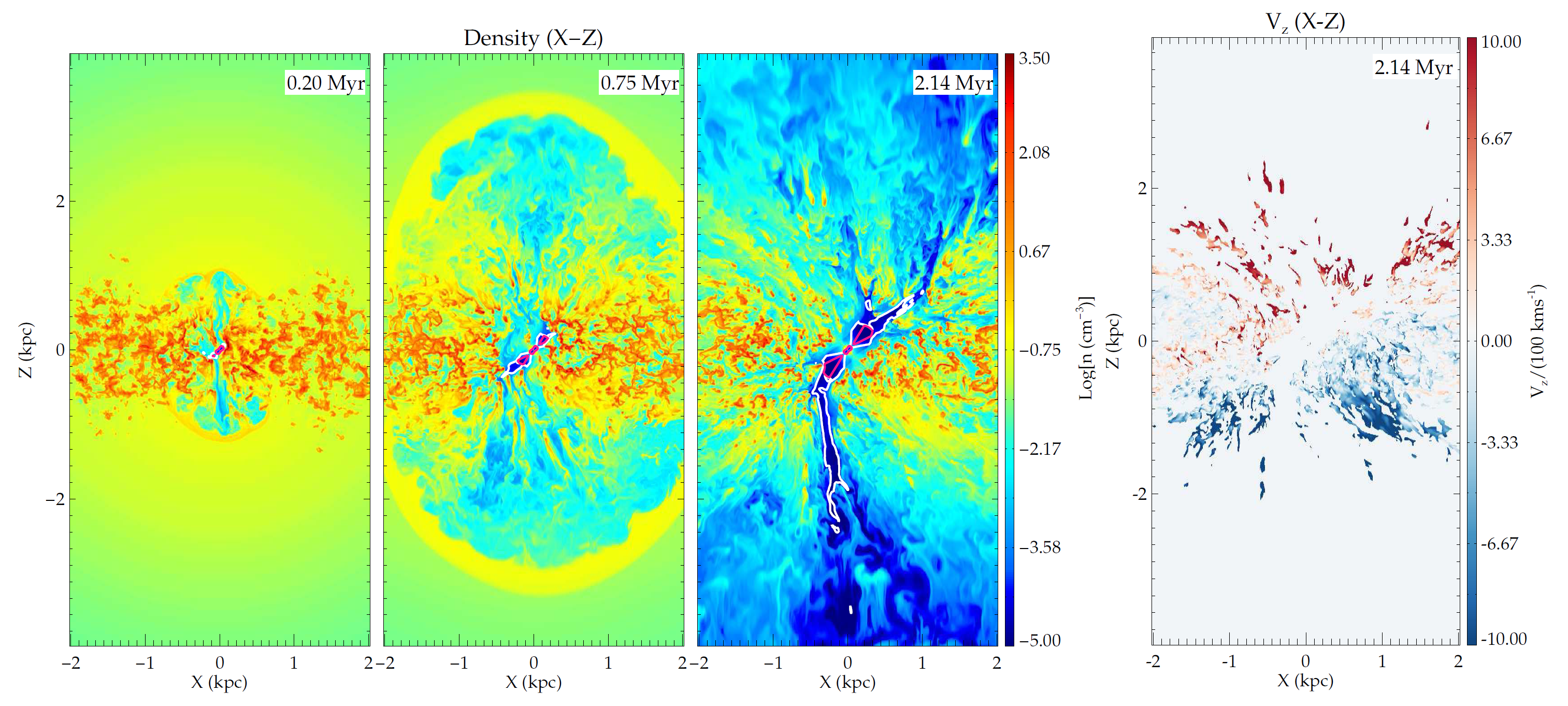}
\caption{One of the simulated interactions between the jet and the gas by~\citet{mukherjee18b}, in~this case within $45^\circ$ inclination between the jet axis and the galaxy disk plane. The left panels display the time evolution of the gas density along the radio axis plane. The~right panel shows the projected velocity of the gas, where a clear, enhanced component is observed perpendicularly to the jet. As~a result of the enhanced interaction with the ISM, in~comparison to scenario with the jets being launched perpendicularly to the disk, the~jet is decelerated and generates higher turbulence dispersion, mainly towards the path of less resistance perpendicularly to~it.}
\label{fig:mukherjee}
\end{figure}

As discussed in~\citet{venturi21}, the~galaxies presenting outflows perpendicular to the radio jet are usually hosts of either compact or low-power radio jets (with some exceptions, like 3C 33), and~thus the bulk of the jet emission is close to the nucleus, within~the inner kpc or less. These jets are also not perpendicular to the galaxy disk, with~low inclination angles between these structures, resulting in a larger impact of the jet in the ISM. Galaxies that present jets oriented perpendicularly to the galaxy disk plane do not display the same behavior. The~spatial correlation of the regions with high gas velocity dispersion with high line ratios related to shocks (as shown in Figure~\ref{fig:3c33}) indicates that these outflows are shock-driven due to the jet interaction with the galaxy disk gas. In~galaxies with large-scale jets, usually FR II objects, with~lobes extending tens of kpcs from the galaxy nucleus, the~jet could have already cleared much of the gas in its path. Additionally, high-power jets ($\gtrsim 10^{45}\,$\ergs) penetrate the clumpy ISM and dig their way out through the gas easier when compared to low-power jets~\citep{wagner11,mukherjee18b}.

A similar scenario has been reproduced with radio-mode feedback models. The~jet power and its inclination relative to the galaxy disk determines how the jet affects the gas in the NLR and ISM, as~discussed in~\citet{mukherjee18b}. Jets inclined closer to the disk influence the gas more than when its orientation is perpendicular to it, generating shock-driven wide-angle outflows along the disk's minor axis and enhancing turbulent dispersion along the disk. As~shown by the gas density evolution along the jet axis in Figure~\ref{fig:mukherjee}, with~a jet axis inclination of $45^\circ$ with the disk plane, the~jet interacts strongly with the gas, and~the denser gas (when compared to higher disk latitudes) deflect the jets, which is decelerated. The direction of least resistance is where the turbulence is propagated, which is along the direction perpendicular to the radio~jet.


\section{Maintenance Mode Feedback---Red~Geysers}
\label{sec:main}

In~\citet{comerford20}, the~authors have identified in the MaNGA/SDSS galaxy survey (Mapping Nearby Galaxies with the Apache Point Observatory of the Sloan Digital Sky Survey)~\citep{bundy15} a number of low-luminosity AGNs in radio using the NVSS and FIRST radio surveys, many of which were not detected in the optical observations. They separated the sample into radio-mode and radio-quiet AGNs and, comparing the two subsamples, found that the radio-mode ones are preferably hosted by early-type galaxies, have older stellar populations, and lower star-formation rates. Such galaxies are sometimes referred to as being ``red and dead,'' and~in~\citet{comerford20} the authors suggest that feedback from radio jets could be the source of the suppression of star~formation.

But the authors also point out that one caveat of the conclusion above is the fact that early-type galaxies have less gas in the inner region, implying low mass accretion rates to the nuclear SMBH, which leads to advection-dominated accretion flows, e.g.~\citep{narayan08,nemmen14}. It is well known that such accretion flows favor the formation of radio jets. Thus, instead of radio jets being the source of feedback that would lead to red and dead galaxies, they are more a consequence of the low accretion rates to their SMBH, resulting from the scarcity of gas in the nuclear region of early-type galaxies. In~addition, while radio activity in AGNs are short-lived phenomena a few million years old, continuum optical and IR galaxy features, as~well as the central stellar populations, are hundreds of millions to billions of years old, which means that the radio feedback has little to none effect in these structures. On~the other hand, there is still mass loss due to the process of stellar evolution that could still lead to some star formation that seems not to be there, and this could indeed be attributed to feedback---even if mild, produced by low-power radio jets, precluding new star formation. It is also important to take into consideration recurrent radio activity where relic emission is observed, mainly in merger remnants such as Centaurus A or NGC 3801. In~these galaxies, the older activity could indeed have had an impact on the past star formation history of the host~galaxy.

\begin{figure}[H]
\hspace{-0.6cm}
\includegraphics[width=\textwidth]{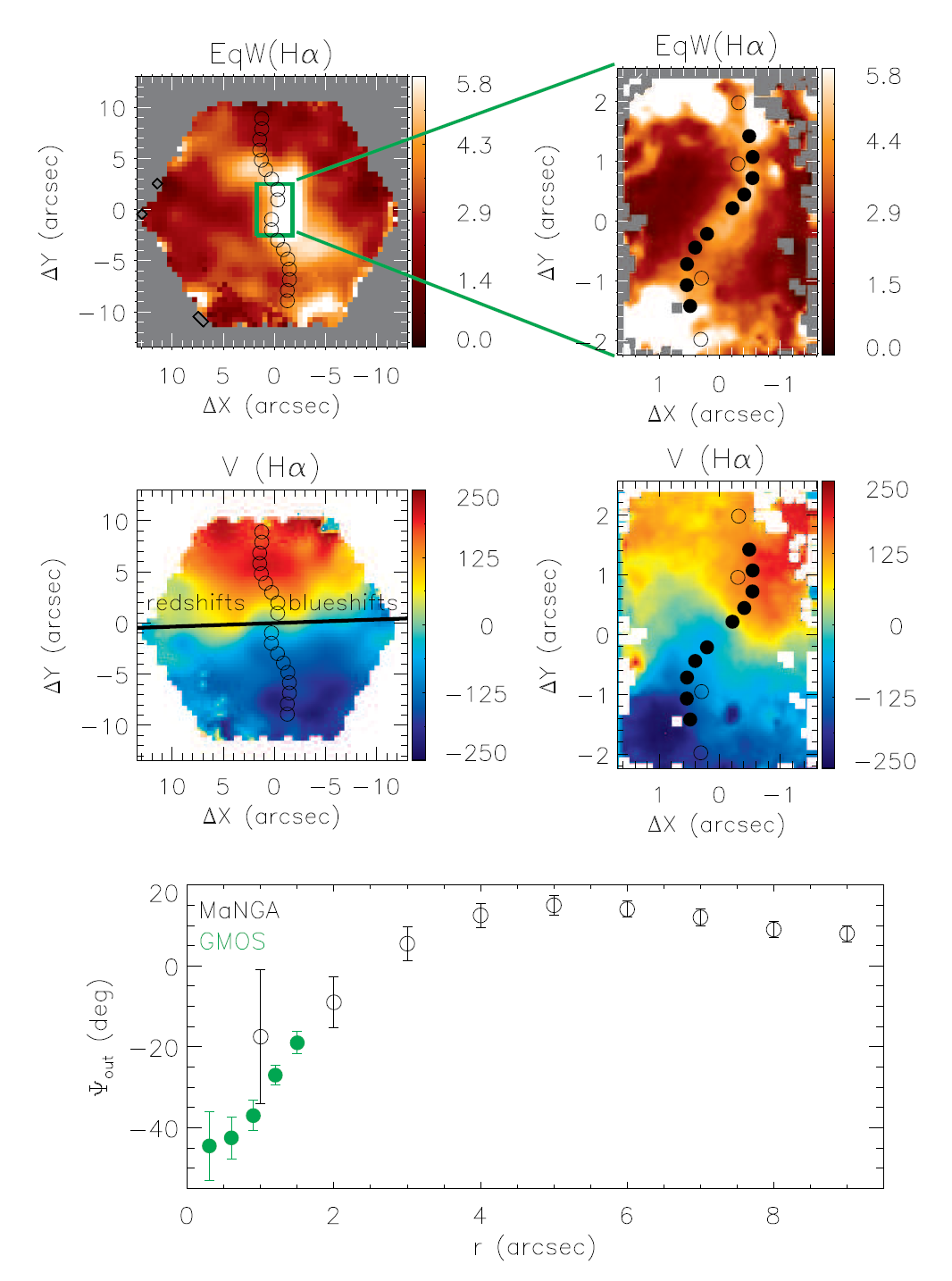}
\caption{Outflow orientation of the prototype Red Geyser Akira Figure 4 in~\citep{riffel19}. SDSS MaNGA (left panels) and Gemini-GMOS IFS (right panels) display different scales of the galaxy (Gemini FoV represented by the red rectangle in the top left panel), with~top and middle panels showing the H$\alpha$ equivalent width and velocity maps, respectively. Circles represent the orientation of outflowing gas after~interpreting that these regions cannot be reproduced by disk-like rotation, where open and closed circles are using MaNGA and GMOS data, respectively. Redshifted and blueshifted velocities obtained from the stellar kinematics are identified along the line of the nodes measured in the stellar velocity field. The~bottom panel displays the outflow orientation relation with distance to the nucleus. The~variation in outflow orientation is interpreted as precessing winds due to the orientation of the accretion disk in relation to the SMBH~spin.}
\label{fig:red_geys}
\end{figure}

Low-power jets can produce a mild outflow, producing what is being called ``maintenance mode feedback,'' also observed in a class of galaxies that have been called Red Geysers, in~which a low-luminosity AGN seems to be the source of large-scale but mild outflows that were discovered in observations with the MaNGA~\citep{cheung16}. The~Red Geysers frequently show a radio-source at the nucleus~\citep{roy21}, and~recent optical IFS of the inner region at a resolution at the galaxies of a few 100 pc using the Gemini telescopes~\citep{ilha22} have revealed nuclear outflows that are misaligned relative to the larger scale outflows discovered in the MaNGA survey. This misalignment has been argued by~\citet{riffel19} as being due to the precession of a nuclear outflow/jet, as~illustrated in Figure~\ref{fig:red_geys}.

Using MaNGA/SDSS datacubes of a sample AGN,~\citet{wylezalek20} have also shown that low-luminosity AGNs can introduce mild kinematic disturbances over most of the galaxy extension as mapped via the $W_{80}$ kinematic parameter, which measures the width of line profiles comprising 80\%\ of their fluxes. Using a control sample, Gatto~et~al. [in preparation] confirms the larger $W_{80}$ in AGNs than controls down to the lowest AGN luminosities, configuring a mild, maintenance mode~feedback.

Another possible scenario for the misalignment between the small and large-scale ionized gas is the radio detection limit from observations using telescopes and arrays such as the VLA. Recent and future radio telescopes such as LOFAR and SKA will explore further the faint emission of older jet feedback, which may correlate with the large-scale ionized gas orientation, while the gas closer to the nucleus is affected by younger and more compact activity. Example cases of these phenomena are NGC 5813~\citep{randall11} and \mbox{Mkr 6~\citep{mingo11}.} A~recent report of LOFAR observations by~\citet{webster21} describes a large sample of intermediate scale radio sources in low luminosity AGNs at 150 MHz. As~these new observations expand the samples of faint sources, we will be able to further constrain how the radio feedback interacts with its galaxy~hosts.


\section{Summary}

AGN radio activity and its relation with the host galaxy has been the subject of detailed studies in the context of galaxy evolution. With~the further development of IFS instruments, we have been able to trace and characterize gas outflows originating in jet--gas interactions in nearby radio galaxies. In~this review, we focused on the discussion of the interplay of radio jets with the circumnuclear gas of radio AGNs, displaying clear examples of where these jets have an impact on the galaxy's ISM. We showed how galaxy interactions seem to have an important role in the triggering radio activity, usually found in early-type galaxies, via the feeding of the nuclear~SMBH. 

Feedback in the form of outflows produced via the interaction of the radio-jet with the circumnuclear gas is observed not only in ionized gas phases, but~also in molecular gas, which should represent the bulk of the gas mass within the outflows, as discussed in the context of AGN scaling relations. Jet--gas coupling and orientation of the outflows are important parameters to understand the produced feedback, as~even low power radio jets can be well coupled with the gas and generate outflows. These outflows are observed both in the direction of the radio jet but also perpendicularly to it, usually traced with enhanced velocity dispersions perpendicular to the radio~jet.

Although there are a few radio AGNs that produce strong feedback on the host galaxies, most radio sources in the near Universe present outflow kinetic powers that do not reach $1\% L_{bol}$, and~thus do not provide a strong and immediate impact on the host galaxy. Instead, they act to heat the ISM gas, preventing star formation, slowing the galaxy mass build-up process, and limiting the stellar mass growth in a ``maintenance mode''~feedback.

The era of the James Webb Space Telescope will allow us to perform similar spatially resolved studies at higher redshifts, revealing important information on AGN feedback across cosmic time. With~the aid of incoming surveys such as 4MOST and WEAVE and the future IFS generation (such as ELT/HARMONI), we will be in a better condition to establish the role of radio feedback on galaxy evolution, not only in high-power jets but also in low radio luminosity AGNs, which also seem to be an important piece of the~puzzle.

\vspace{6pt} 



\authorcontributions{All authors have read and agreed to the published version of the manuscript.} 

\funding{This research received no external funding}

\dataavailability{The data underlying this article will be shared on reasonable request to the authors.}  

\acknowledgments{The authors thank the anonymous referees for their valuable contributions on improving this review.}

\conflictsofinterest{The authors declare no conflict of interest.}

\begin{adjustwidth}{-\extralength}{0cm}

\reftitle{References}

\end{adjustwidth}
\end{document}